\documentclass[amsmath,amssymb,aps,twocolumn,nofootinbib,prd]{revtex4-2}

\usepackage{graphicx}
\usepackage{dcolumn}
\usepackage{bm}
\usepackage[dvipsnames]{xcolor}
\usepackage[normalem]{ulem}
\usepackage{enumerate}

\newcommand{\aap}{Astron.\ Astrophys.}
\newcommand{\mnras}{Mon.\ Not.\ R.\ Astron.\ Soc.}

\newcommand{\apjl}{Astrophys.\ J.\ Lett.}
\newcommand{\apjs}{Astrophys.\ J.\ Suppl.\ Ser.}
\newcommand{\araa}{Ann.\ Rev.\ Astron.\ Astrophys.}

\newcommand{\jcap}{J. Cosmology \& Astropaticles.}

\newcommand{\Eqgi}[1]{E_{\text{LIV}#1}}
\newcommand{\Eqg}{E_{\text{LIV}}}
\newcommand{\Epl}{E_\text{Pl}}
\newcommand{\diff}{\mathrm{d}}
\newcommand{\Ecpt}{E_\gamma^\text{(Cpt)}}
\newcommand{\tcpt}{t_\text{(Cpt)}}
\newcommand{\zgrb}{z_\text{GRB}}

\begin{document}
 

\title[Hint at Non-linear LIV]{The 300 TeV photon from GRB 221009A: a Hint at Non-linear Lorentz Invariance Violation?}%

\author{Dmitry D. Ofengeim}\email{Corresponding Author: ddofengeim@gmail.com}
\author{Tsvi Piran}%

\affiliation{Racah Institute of Physics, The Hebrew University, Jerusalem 91904, Israel}

\date{\today}

\begin{abstract}
The air shower array Carpet-3 detected a 300 TeV photon from the direction of GRB~221009A at 4536 s after the \textit{Fermi}-GBM trigger for this event. If the association with this gamma-ray burst is real, it poses two puzzles. First, why was this photon not absorbed by the extragalactic background light? ``New physics'' beyond the Standard Model is required to explain how it managed to reach  Earth from a cosmological distance. Second, why was this photon detected when the VHE afterglow observed by LHAASO already faded? A novel astrophysical mechanism is required to explain this delay. In this work we show that Lorentz invariance violation (LIV), which arises as a low-energy limit of certain quantum gravity theories, can solve both puzzles. It shifts thresholds of particle interaction and changes the opacity of the extragalactic background, and causes energy-dependent variations of the photon velocity, which changes the photon time of flight. We investigate the LIV parameter space assuming that the 300 TeV photon is a part of the VHE afterglow detected by LHAASO in the TeV range. We identify viable solutions and place stringent two-sided constraints on the LIV energy scale required to resolve the observational puzzles. First-order LIV appears to be incompatible with the constraints set by analyzing the TeV afterglow of this GRB. Viable solutions emerge for higher orders. In particular, the commonly studied second-order subluminal LIV with $\Eqgi{2} = 1.30_{-0.35}^{+0.56} \times 10^{-7} \Epl$ (95.4\%~credibility level; $\Epl$ is the Planck energy) is consistent with all the observed data.
\end{abstract}

\maketitle

\section{Introduction}
\label{sec:intro}

GRB~221009A  was the strongest and brightest gamma-ray burst (GRB) of all time \cite{BOAT}. Its unique observations opened up new windows on GRB phenomena. In addition, the observed TeV emission enables us to explore physics beyond the Standard Model. 

Initial potential claims of the observation of an 18 TeV photon by LHAASO \cite{LHAASO_18TeV_prelim} led to a wave of different beyond-the-standard-model explanations. As GRB 221009A is located at $\zgrb = 0.151$, such a photon would have been absorbed by the extragalactic background light (EBL) due to annihilation of the TeV photon by an infrared one, $\gamma\gamma \rightarrow e^+e^-$ (see e.g. \cite{FranceschiniU2021}). However, a careful analysis of the data~\cite{LHAASOSci2023_13TeV} revealed that this photon's energy is ``only'' $13\,$TeV, relaxing the need for any new physics explanation. On the contrary, observations of the TeV emission from this event led to new strong upper time of flight (TOF) limits on Lorentz invariance violation (LIV), a phenomena in which the velocity of very high energy photons differ slightly from the low-energy speed of light \cite{PiranOfengeimPRD2024_LIV,LHAASOSci2024_LIV}. 
It is considered as a possible low-energy manifestation of  quantum gravity.

The air shower array Carpet-3 \cite{Carpet-3_device} detected a photon of energy $\Ecpt = 300_{-38}^{+43}\,$TeV from the direction towards GRB~221009A at $\tcpt = 4536\,$s \cite{Carpet3_300TeV} after the Fermi-GBM trigger~\cite{Fermi-GBM}. Presence of other near-PeV sources within the confidence region in the sky is considered, and based on the angular distribution alone there is a 99.1\% probability that this photon is associated with GRB~221009A. This is the first time ever sub-PeV emission is claimed for a GRB. If the association is real, then such a high-energy photon would have been absorbed by EBL \cite{FranceschiniU2021}. Its arrival to Earth cannot be explained within Standard Model physics.

A possible explanation for the ability of such a high energy photon to traverse the EBL is a shift of the threshold for the $\gamma\gamma \rightarrow e^+ e^-$ reaction  due to LIV \cite{Gonzalez-Mestres1997,Coleman1999,Kifune1999,Aloisio+PRD2000,Amelino-Camelia2001}. In the simplest and most commonly discussed version of LIV the dominant effect at low energies is a modification of the particle's dispersion relation that is determined by just two parameters, the power $n$ in which LIV arises at low energies, and $\Eqg$ the corresponding energy scale:
\begin{equation}
    \label{eq:LIVdef}
    E^2 = m^2 c^4 + p^2 c^2 \left[ 1 - \left( \frac{E}{\Eqg} \right)^n \right].
\end{equation}
Here $m$ is the rest mass of the relevant particle and the fundamental constant $c$ means the speed of massless particles, say photons, in the limit $E/\Eqg \to 0$. Two immediate implications of this modification are shifts of reaction thresholds (in particular of the threshold for $\gamma\gamma \rightarrow e^+ e^-$) and an energy-dependent time of flight of high-energy photons \cite{Amelino+Nat1998,Adazzi+PrPNP2022}. In principle, LIV may make the speed of high-energy photons both lower or greater than $c$, but only subluminal LIV, written in Eq.~\eqref{eq:LIVdef}, shifts the threshold of $\gamma\gamma$ reaction to higher values and suppresses the EBL absorption. Following the discovery of the Carpet-3 photon 
Galanti and Roncadelli~\cite{GalantiRoncadelli2025} suggested that this photon managed to traverse the EBL due to an LIV-induced threshold shift. They show that for linear LIV, $n=1$, the quantum gravity energy scale $\Eqgi{1}$, required for this photon to survive, should be less than or about $3\times 10^{20}\,\text{GeV} \sim 25\Epl$. 
The Planck energy $\Epl = 1.22\times 10^{19}\,$GeV is a natural scale to expect the 
quantum gravity effects to come into play.


However, at the time $\tcpt$ the TeV emission from GRB 221009A diminished \cite{LHAASOSci2023_aftrglw} and only low-energy signals were detected (see e.g. \cite{Tak2025}). This raises the question whether the Carpet-3 photon can be associated with the GRB. Such an association requires a mechanism, either intrinsic or acting while traversing the Universe, that explains the more than one hour delay of this photon with respect to the earlier TeV signal. In principle, LIV in the form of Eq.~\eqref{eq:LIVdef} gives rise to a time-of-flight delay. But linear LIV with $\Eqgi{1} = 25\Epl$ results in a LIV TOF delay of only about $60\,$s. This is insufficient to relax the tension in the arrival time. Is there an LIV explanation consistent with emission of the 300\,TeV photon during the main TeV flare? In this work, we gather all TOF and EBL constraints on LIV that follow from GRB~221009A observations and search for a consistent solution.

\section{Timing of the Carpet-3 photon}
\label{sec:when}

An intense VHE afterglow of GRB~221009A was detected by LHAASO in the range $0.2...13\,$TeV \cite{LHAASOSci2023_aftrglw,LHAASOSci2023_13TeV}. The first photons of more than 60000 in total were registered at $t_0 = 226\,$s after the \textit{Fermi}-GBM trigger \cite{Fermi-GBM} for the prompt emission of this burst, and totally faded away within 3000\,s. The Carpet-3 photon arrived $\tcpt = 4536\,$s after the GBM trigger and $\tcpt - t_0 = 4310\,$s after the beginning of the TeV afterglow. Since association of this photon with the GRB would require some physics beyond the Standard Model (otherwise such a photon could not have traversed the EBL), one should explore carefully the evidence for coincidence as well as  physical processes that could allow for such a delayed emission. 

The energy fluence implied by this photon is consistent with extrapolation of the LHAASO TeV spectrum (see figure~7 of \cite{Carpet3_300TeV}). This  supports the association. However, this photon was observed at a time that the LHAASO TeV afterglow diminished a factor of at least $10^{-3.5}$ below the peak. There are two general options to explain this. Either a mechanism that produced this photon is different from the one that produced the afterglow, or the photon was emitted together with the rest of the TeV emission but experienced a time-of-flight delay---e.g., as a result of LIV. We begin by exploring the former option.  Then we explore the implication that the LHAASO TeV afterglow and the Carpet-3 300\,TeV photon were produced by the same mechanism.

\subsection{Possibility of extremely large intrinsic delay}
\label{sec:when:largedelay}

Is there an intrinsic mechanism that could produce the delayed arrival of such a photon? The authors of the observational paper \cite{Carpet3_300TeV} 
discarded several options: inverse Compton, proton synchrotron, and a photohadronic scenario. All these mechanisms were ruled out just because of their inability to produce efficiently a 300 TeV photon, regardless of the possibility of emission delay. To solve the puzzle, the authors \cite{Carpet3_300TeV} proposed the so called neutron beam model \cite{DermerAtoyan2004}. 

In this scenario, PeV neutrons escape from the source during the prompt phase of the GRBs. These neutrons interact with interstellar matter surrounding the GRB and produce ultra-relativistic electron-positron pairs. These pairs produce, in turn,  multi-TeV photons via synchrotron radiation on the surrounding magnetic field. The time delay arose in this model due to off-axis spreading of the neutron beam. While this model has several theoretical drawbacks (see Appendix~\ref{sec:app}), we focus here on a major observational contradiction. An essential and inevitable component of this model \cite{DermerAtoyan2004} is the production, during the interaction of the neutrons and the interstellar matter, of a comparable flux of multi-TeV neutrinos. However, 90\% upper limits from IceCube observations \cite{Icecube2023,Kruiswijk2023} are at least an order of magnitude below the Carpet-3 fluence. 

Lacking a clear possible mechanism for a delayed emission of the 300 TeV photon we continue in the rest of our discussion assuming that this component belongs to the same VHE afterglow as detected by LHAASO.

\subsection{Temporal characteristics of GRB 221009A VHE afterglow}
\label{sec:when:temporal}

\begin{figure*}
    \centering
    \includegraphics[width=\textwidth]{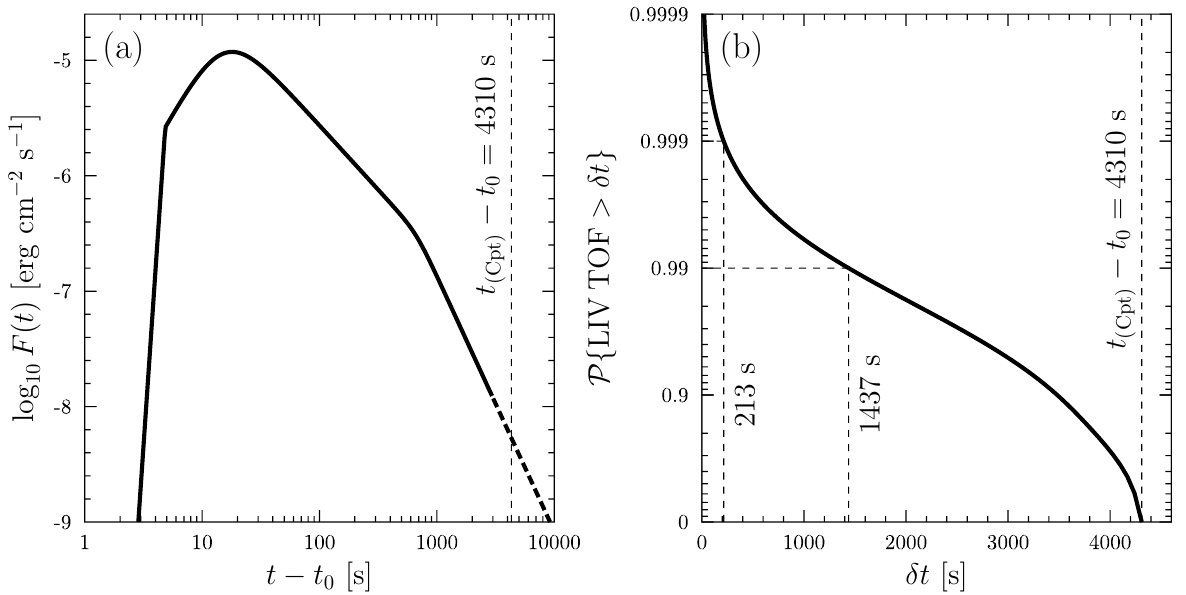}
    \caption{\label{fig:temporal} (a) The best-fit light curve of the GRB~221009A afterglow measured by LHAASO in the $0.3...5\,$TeV range \cite{LHAASOSci2023_aftrglw}. The dashed part depicts the times when no reliable signal were detected by LHAASO. (b) The probability distribution Eq.~\eqref{eq:prob_LIVTOF} for the LIV TOF delay of the Carpet-3 photon, estimated from the LHAASO afterglow light curve.}
\end{figure*}

Let us assume (without discussing a specific emission mechanism) that this 300\,TeV photon came from the same processes as the photons detected by LHAASO in the range $0.2...13\,$TeV \cite{LHAASOSci2023_aftrglw,LHAASOSci2023_13TeV}. 
The observed LHAASO spectrum does not show a significant spectral evolution with time: the spectral slope varies from $-2.5$ to $-2.2$ during the whole LHAASO observation that lasted about 3000\,s. 
If this property spans over the whole range from sub-TeV to sub-PeV energies, the 300 TeV light curve should have the same shape as the the one observed by LHAASO. Integrating the light curve, we can estimate the probability for the Carpet-3 photon to be emitted at a given moment of time.

The LHAASO light curve $F(t)$ is shown in Fig.~\ref{fig:temporal}(a). 
{It begins} at $t_0 = 226\,$s after the Fermi-GBM trigger~\cite{Fermi-GBM}. Treating the light curve as a probability distribution for the emission moment, we can write the cumulative probability that our photon was emitted before $t$ as 
\begin{equation}
    \label{eq:prob_before_t}
    \mathcal{P}\{\text{emitted before } t\} = \int_{t_0}^t F(t') \diff t' \bigg/ \left( \int_{t_0}^{\tcpt} F(t') \diff t' \right).
\end{equation}
If the photon was emitted at $t$, then its LIV TOF delay is $\delta t = \tcpt - t$. It vanishes
\footnote{LIV TOF cannot be negative since the latter corresponds to superluminal LIV, which is inconsistent with  suppressed, rather than enhanced, EBL absorption.}
for the latest possible emission moment $\tcpt$, and its maximal value is $\tcpt - t_0 = 4310\,$s. The probability distribution for LIV TOF is
\begin{equation}
    \label{eq:prob_LIVTOF}
    \mathcal{P}\{\text{LIV TOF} > \delta t\} = \mathcal{P}\{ \text{emitted before } \tcpt - \delta t \}.
\end{equation}
This probability distribution is shown in Fig.~\ref{fig:temporal}(b). Since the afterglow is very faint after $t\gtrsim 3000\,$s, small LIV TOF delays are strongly disfavored. For instance, at $99\%$ credibility the required LIV TOF is larger than $1437\,$s, and at $99.9\%$ credibility it is larger than $213\,$s.

\section{GRB  221009A LIV constraints}
\label{sec:LIV}

We turn now to consider different constraints that can be set from GRB 221009A on the LIV parameters $\Eqg$ and $n$.

\subsection{Time-of-flight delays}
\label{sec:LIV:TOF}

The standard formula for LIV TOF in  standard cosmology is \cite{JacobPiranJCAP2008}
\begin{subequations}
\label{eq:dt_LIV_gen}
\begin{gather}
\begin{split}   
    \delta t^\text{(LIV)}(E_\gamma&,z {; \Eqg, n})  
    \\
    = \frac{n+1}{2H_0}& \left( \frac{E_\gamma}{\Eqg} \right)^n \int_0^z \frac{(1+\zeta)^n \diff\zeta}{\sqrt{\Omega_m(1+\zeta)^3 + \Omega_\Lambda}} 
    \\
    &= \frac{1}{H_0} \left( \frac{E_\gamma}{\Eqg} \right)^n [f(1+z,n) - f(1,n)],
\end{split}
\\
\begin{split}
    f(x,n) = \frac{n+1}{2n-1}&\frac{x^{n-1/2}}{\sqrt{\Omega_m}} 
    \\
    \times {}_2F_1 &\left( \frac{1}{2}, \frac{1}{6}-\frac{n}{3}, \frac{7}{6}-\frac{n}{3}, -\frac{\Omega_\Lambda/\Omega_m}{x^3} \right).
\end{split}
\end{gather}
\end{subequations}
In this work, we use $H_0 = 67.4\,\text{km}\:\text{s}^{-1}\,\text{Mpc}^{-1}$, $\Omega_m = 0.315$ and $\Omega_\Lambda = 1 - \Omega_m$ \cite{PLANCK_final}. The LHAASO afterglow gives a stringent constraint on $\Eqg$ from below. In \cite{PiranOfengeimPRD2024_LIV,LHAASOSci2024_LIV} it is given for $n=1$ and $n=2$. Both these constraints roughly correspond to the condition $\delta t^\text{(LIV)} < 3.5\,$s. The energy range used in \cite{PiranOfengeimPRD2024_LIV,LHAASOSci2024_LIV} is $0.3...5\,$TeV, so we can extend these constraints to arbitrary $n$ as 
\begin{equation}
    \label{eq:dt_LHAASO}
    \delta t^\text{(LIV)}(E_\gamma, \zgrb {; \Eqg, n})\Bigr|_{E_\gamma = 0.3\,\text{TeV}}^{E_\gamma = 5\,\text{TeV}} < 3.5\,\text{s}.
\end{equation}
This limit is plotted in Fig.~\ref{fig:n-lgEqg} by the blue lines.

The Carpet-3 photon clearly satisfies 
\begin{equation}
    \label{eq:dt_Carpet_bottom}
    \delta t^\text{(LIV)}(\Ecpt, \zgrb {; \Eqg, n}) < \tcpt - t_0 = 4310\,\text{s}.
\end{equation}
This condition is drawn in Fig.~\ref{fig:n-lgEqg} by the green bands. Their thickness corresponds to the 68\%\ range of $\Ecpt$ from $262\,$TeV to $343\,$TeV. From the probabilistic treatment of the light curve (Sec.~\ref{sec:when:temporal}), we can say that, at correspondent credibility levels,
\begin{equation}
    \label{eq:dt_Carpet_99}
    \delta t^\text{(LIV)}(\Ecpt, \zgrb {; \Eqg, n}) > 1437\,\text{s} \quad (99\%)
\end{equation}
and 
\begin{equation}
    \label{eq:dt_Carpet_999}
    \delta t^\text{(LIV)}(\Ecpt, \zgrb {; \Eqg, n}) > 213\,\text{s} \quad (99.9\%).
\end{equation}
These conditions are shown in Fig.~\ref{fig:n-lgEqg} by the orange and yellow bands (again, their thickness corresponds to 68\%\ uncertainty of $\Ecpt$).

\subsection{EBL absorption suppression}
\label{sec:LIV:EBL}

Within the Standard Model, the threshold energy of a low-energy photon that can interact with
a VHE one
with energy $E_\gamma(1+z) $ at the epoch of scattering is $\epsilon_\text{th} = m_e^2 c^4/[E_\gamma(1+z)]$. For VHE photons the low-energy 
{target} photons are in the infrared to near-microwave range.
The standard optical depth $\tau(E_\gamma, z)$ with respect to this process is extracted from observations of multiple high-energy sources as well as from galaxy counts and estimates of emission at this energy band. In this work we use the results by \cite{Saldana-Lopez+MNRAS2021} with additional constraints given by LHAASO observations~\cite{LHAASOSci2023_13TeV}. In particular, for the source at $z=0.151$ this interaction would have suppress the spectrum above a  few TeV  and would make 
{the} EBL completely opaque for $E_\gamma \gtrsim 15\,$TeV.

Subluminal modification of the dispersion relation due to LIV shifts the threshold to higher energies. The cross section peaks around the threshold, and as typically the number of target photons decreases with their energy, an increase in the threshold lowers the optical depth and suppresses the EBL absorption. 
If the dominant low-energy LIV manifestation is the change in the dispersion relation expressed in Eq.~\eqref{eq:LIVdef}  then the only LIV effect on the EBL absorption is this modification of $\epsilon_\text{th}$ \cite{JacobPiranPRD2008_EBL}: 
\begin{equation}
    \label{eq:thres_LIV}
    \epsilon_\text{th}^\text{(LIV)} = \frac{m_e^2 c^4}{E_\gamma (1+z)} + \frac{1-1/2^n}{4}\frac{E_\gamma^{n+1}(1+z)^{n+1}}{\Eqg^n}.
\end{equation}
For $z \ll 1$ one can calculate\footnote{For $z=\zgrb=0.151$ this approximation is accurate enough for calculating $\Eqg-n$ relation at a fixed $\tau^\text{(LIV)}$.} the modified optical depth $\tau^\text{LIV}$ as the standard (that is with no LIV) $\tau$, by replacing $E_\gamma$ with the effective energy 
$E_\gamma^\text{(eff)} \equiv m_e^2 c^4 / [\epsilon_\text{th}^\text{(LIV)} (1+z)]$. Thus, 
\begin{subequations}
\label{eq:tau_LIV}
\begin{align}
    &\tau^\text{(LIV)}(E_\gamma,z {; \Eqg, n}) \approx \tau\left(E_\gamma^\text{(eff)}, z \right),
    \\
    &E_\gamma^\text{(eff)} = 
    \frac{E_\gamma}{ 1 + \frac{1-1/2^n}{4}\frac{E_\gamma^{n+2}}{m_e^2 c^4 \Eqg^n}(1+z)^{n+2} }
\end{align}
\end{subequations}
Notice that this function increases with $\Eqg$ and decreases with $n$.

Analysis of GRB~221009A with respect to the EBL absorption gives two constraints on LIV. The first one arises from the LHAASO observation of photons with energies up to $13\,$TeV \cite{LHAASOSci2023_13TeV}, which are consistent with the standard EBL absorption. This gives an upper limit on the quantum gravity scale, estimated in \cite{LHAASOSci2023_13TeV} for {$n=1$} as $\Eqgi{1} \gtrsim 1.5\Epl$. With Eq.~\eqref{eq:tau_LIV} this transforms to $\tau^\text{(LIV)} \gtrsim {6.7}$. Thus, for arbitrary $n$ we take this constraint as
\begin{equation}
    \label{eq:EBL_LHAASO}
    \tau^\text{(LIV)}(13\,\text{TeV},\zgrb {; \Eqg, n}) > {6.7}.
\end{equation}
This condition is plotted in Fig.~\ref{fig:n-lgEqg} by the black dashed lines. It is weaker than the constraint Eq.~\eqref{eq:dt_LHAASO} due to TOF analysis of the LHAASO afterglow. 

The second constraint arises from the 300\,TeV photon observed by Carpet-3 and was not absorbed by the EBL. For $n=1$ it implies $\Eqgi{1} \lesssim 3\times 10^{20}\,\text{GeV} \approx 25\Epl$ \cite{GalantiRoncadelli2025}. With Eq.~\eqref{eq:tau_LIV} this gives $\tau^\text{(LIV)} \lesssim 3.3$. We extrapolate this result to arbitrary $n$ as
\begin{equation}
    \label{eq:EBL_Carpet}
    \tau^\text{(LIV)}(\Ecpt,\zgrb {; \Eqg, n}) < 3.3.
\end{equation}
This upper boundary is shown in Fig.~\ref{fig:n-lgEqg} by the solid black line, where the 
{68\%\ credible} interval for $\Ecpt$ is taken into account.

\subsection{Combined constraints on LIV parameters}
\label{sec:LIV:gather}

\begin{figure*}
    \centering
    \includegraphics[width=\textwidth]{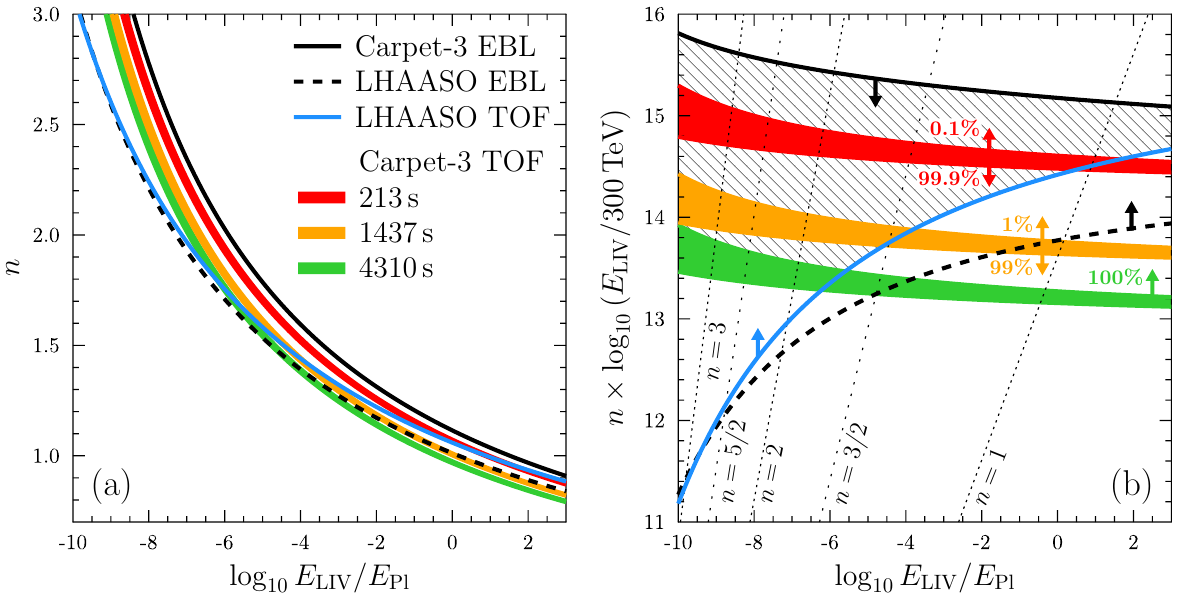}
    \caption{Constraints on the LIV parameters $\Eqg$ and $n$ arising from analysis of GRB~221009A's afterglow assuming that the Carpet-3 photon is associated with this event. Panel (b) is the same as (a) but the main trend of the lines $n\propto 1/(\text{const} + \log \Eqg)$ is subtracted. The thick black line is the constraint that the 300\,TeV photon is not absorbed by EBL (Eq.~\ref{eq:EBL_Carpet}). The black dashed line is the EBL absorption constraint from the LHAASO afterglow observation (Eq.~\ref{eq:EBL_LHAASO}). The blue line is the constraint from the TOF delay analysis of the LHAASO afterglow (Eq.~\ref{eq:dt_LHAASO}). The green, orange, and red bands stand for the conditions that the LIV TOF delay of the 300\,TeV photon is 213\,s (Eq.~\ref{eq:dt_Carpet_bottom}), 1437\,s (Eq.~\ref{eq:dt_Carpet_99}), and 4310\,s (Eq.~\ref{eq:dt_Carpet_999}), correspondingly (bands thickness show the 68\%\ credible range for $\Ecpt$). The percents near the arrows on the panel~(b) show the probabilities that the LIV TOF delay is less or greater than the given value according to the association with the GRB afterglow, Eq.~\eqref{eq:prob_LIVTOF}. The single-hatched region on the right is the most conservative inference on the LIV parameters. Within this range LIV TOF 
    is consistent with the LHAASO data
    \cite{PiranOfengeimPRD2024_LIV,LHAASOSci2024_LIV} 
    and it satisfies the condition $\delta t^\text{(LIV)} < \tcpt-t_0$, and LIV threshold shifts enable the Carpet-3 photon to traverse the EBL. The assumption that the 300 TeV light curve is the same as the lower energy LHAASO light curve yields much stricter limits, as depicted by the red and green stripes.}
    \label{fig:n-lgEqg}
\end{figure*}

Figure~\ref{fig:n-lgEqg}(a) collects together all the constraints discussed in Secs.~\ref{sec:LIV:TOF} and~\ref{sec:LIV:EBL}. The lines almost merge, but they have a common trend. As one can see from Eqs.~\eqref{eq:dt_LIV_gen}---\eqref{eq:EBL_Carpet}, 
{the boundaries of all these constraints}
can be roughly described as $(E_\gamma/\Eqg)^n \sim \text{const}$. This implies, in turn, $n \propto 1/(\text{const} + \log\Eqg)$. The $\text{const}$ value depends on the chosen photon energy and on the corresponding limiting value for a given constraint. In Fig.~\ref{fig:n-lgEqg}(b) we take the former as 300\,TeV and subtract this trend from all the curves in the panel~(a). We also show in Fig.~\ref{fig:n-lgEqg}(b) several lines of constant $n$. 

Even regardless of the distribution~\eqref{eq:prob_LIVTOF}, the association of the 300\,TeV Carpet-3 photon with GRB~221009A puts two-sided constraints on the quantum gravity scale $\Eqg$ for each value of $n$:  from above due to the EBL absorption, Eq.~\eqref{eq:EBL_Carpet} corresponding to the solid black line in Fig.~ \ref{fig:n-lgEqg}, and from below due to TOF analysis, Eqs.~\eqref{eq:dt_LHAASO} and~\eqref{eq:dt_Carpet_bottom} corresponding to the green and blue lines in this figure. The corresponding region in Fig.~\ref{fig:n-lgEqg}(b) is  single-hatched. Notice that it also satisfies the condition that the EBL absorption for the LHAASO afterglow is not affected by LIV [dashed line, Eq.~\eqref{eq:EBL_LHAASO}], which is weaker than the TOF analysis of those data.

However, if one assumes, following the discussion in Sec.~\ref{sec:when:temporal}, that the afterglow has the same light curves at a few TeV and at 300\,TeV, then the LIV TOF delay for the latter photon has the probability distribution~\eqref{eq:prob_LIVTOF}. In this case we obtain much more stringent constraints. Now the upper limit arises from the distribution of the emission time and it depends on the choice of the credibility level. Specifically in Fig.~\ref{fig:n-lgEqg} this is shown as orange and red stripes for the 99\% and 99.9\% levels, correspondingly. Both these limits are stronger than the constraint  from non-absorption of the 300\,TeV photon by EBL Eq.~\eqref{eq:EBL_Carpet}. The lower limit remains the same as discussed above

For some LIV models this assumption on the 300\,TeV light curve is inconsistent with the LHAASO TOF constraint Eq.~\eqref{eq:dt_LHAASO}. In particular, for the frequently discussed $n=1$ (e.g. \cite{Adazzi+PrPNP2022,Amelino-CameliaSymm2010}) these two constraints are in strong contradiction. They  are consistent with each other just at 0.1\% credibility. For $n=3/2$ LIV, that is inspired by recent considerations of fractional quantum gravity effects \cite[e.g.][]{VaraoLoboBezerraEPL2024}, the constraints are only marginally consistent with each other at $\sim 1\%$ level. 

Self-consistent LIV solutions are possible for higher values of $n$. For such $n$, we find that very roughly 
{$\Eqg \approx 1.3 \times 10^{-7 + 6.6(2/n-1)}\Epl$}. More rigorously we take into account both the $\delta t^\text{(LIV)}$ distribution~\eqref{eq:prob_LIVTOF} and the normal distribution of $\Ecpt$ uncertainties (see Fig.~\ref{fig:n-lgEqg}). For quadratic LIV, which is often considered within the Standard Model Extension approach to LIV \cite{SME2009}, we find acceptable solutions for $\Eqgi{2} = 1.30_{-0.35(-0.52)}^{+0.56(+3.48)}\times 10^{-7}\Epl$ at 95.4(99.7)\%\ credibility level. For cubic and quartic LIV the limits are:  $\Eqgi{3} = 8.5_{-2.2(-3.4)}^{+2.9(+11.9)}\times 10^{-10}\Epl$, and  $\Eqgi{4} = 6.7_{-1.8(-2.7)}^{+2.1(+6.4)}\times 10^{-11}\Epl$ at 95.4(99.7)\% credibility. These numbers are consistent with existing limits on subliminal non-linear LIV~\cite{SME2009,SatuninEPJC2019}.
Higher values of $n$ are unlikely as the implied $\Eqg$ range will be too low to be compatible with other limits.

\section{GRB+LIV, just a coincidence, or something else?}
\label{sec:disc}

\begin{figure*}
    \centering
    \includegraphics[width=0.75\textwidth]{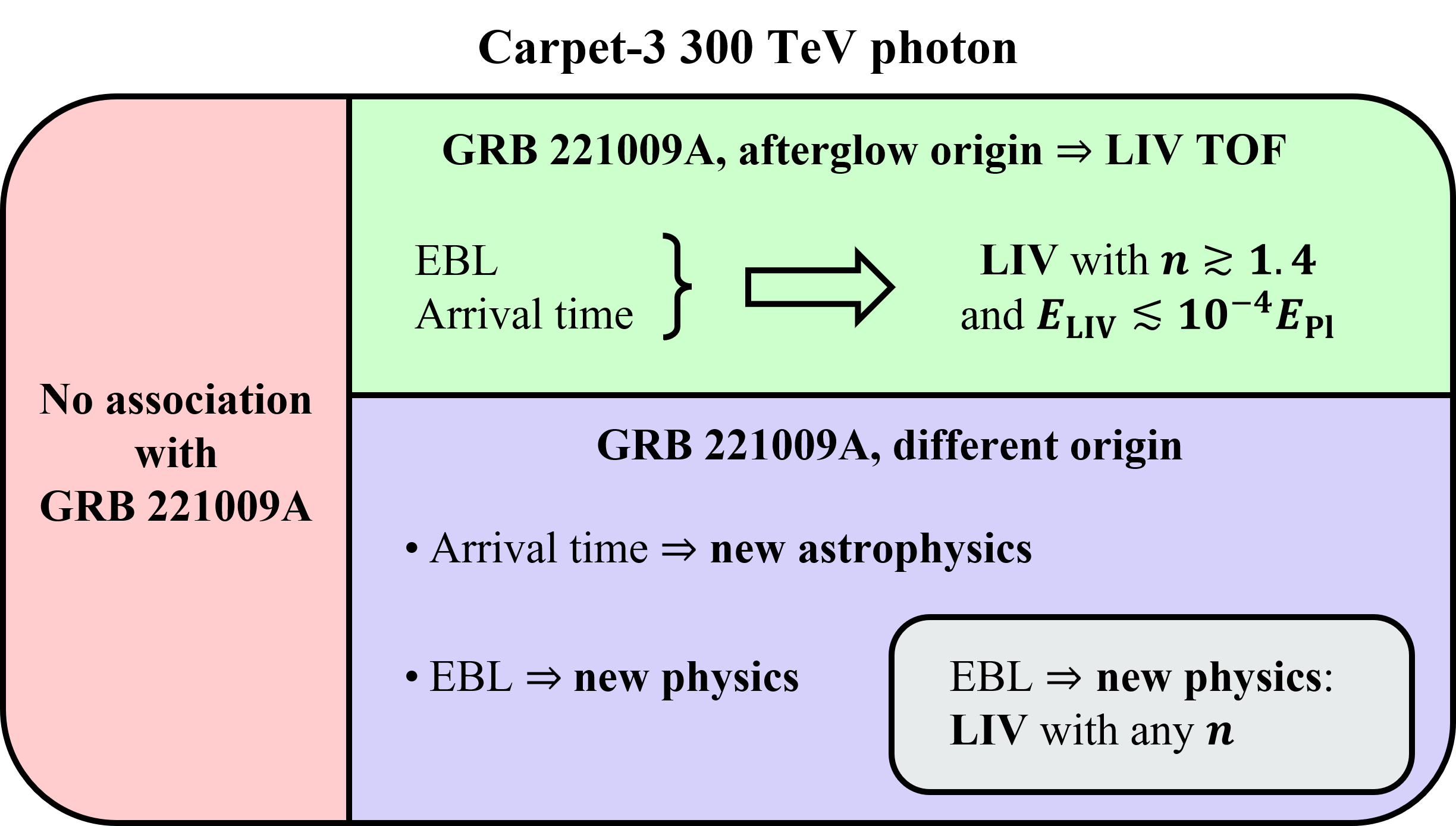}
    \caption{A schematic diagram of the possible solutions to the Carpet-3 photon puzzles. If the photon was not produced by the regular afterglow, this requires both ``new physics'' to explain the suppressed EBL absorption and ``new astrophysics'' to explain the late arrival time. Alternatively if the photon was emitted by the afterglow early on and was delayed on flight, LIV can explain both puzzles.}
    \label{fig:scheme}
\end{figure*}

As we said above, there are two puzzles associated with the observations of the Carpet-3 photon. First its ability to traverse the EBL. Second its late appearance relative to the rest of the VHE signal of this GRB. It is also possible, of course, that the association with GRB 221009A was a coincidence from any other source but GRB~221009A, as $p_\text{chance} = 0.9\%$~\cite{Carpet3_300TeV}.

Any explanation of the EBL puzzle surely requires ``new physics." This can be, for example, some new neutral particle that can traverse the EBL uninterrupted and emit a photon close to Earth, LIV as suggested in \cite{GalantiRoncadelli2025} and discussed earlier, or something else. 

As already mentioned in Sec.~\ref{sec:when:largedelay}, there is no known mechanism according to which a GRB afterglow can produce a detectable signal at this energy with such a late delay. Still, one cannot discard  the possibility that the Carpet-3 photon is indeed related to GRB~221009A, but its origin is different from the LHAASO afterglow. Hence there are two options. The first is that the Carpet-3 photon was not produced by the regular afterglow mechanism.  This option requires a novel mechanism for the  late emission of this photon or models like a revised version of \cite{DermerAtoyan2004}  which result in a late arrival of the photon. All these solutions require ``new astrophysics.'' Alternatively, if the photon was emitted early on by the regular afterglow it must have been delayed ``on route.'' This requires the speed of the Carpet-3 photon to be less than $c$, that is LIV. Overall we can divide the phase space of solutions as described by a diagram shown in Fig.~\ref{fig:scheme}.

Considering the implication for LIV we note that if the Carpet-3 photon did not arise from the afterglow, then the distribution~\eqref{eq:prob_LIVTOF} is irrelevant, and the LIV TOF delay is not constrained from below. Then only the most conservative constraints on $\Eqg$ and $n$ (single-hatched area in Fig.~\ref{fig:n-lgEqg}) are relevant, and low values of $n$, in particular $n=1$, are not excluded. For $n=1$  this implies $9 \Epl  \lesssim \Eqgi{1} \lesssim 25 \Epl$, where the lower limit arises from TOF analysis of the LHAASO afterglow~\cite{PiranOfengeimPRD2024_LIV,LHAASOSci2024_LIV}. 

Carpet-3 collaboration estimates the coincidence probability, i.e. a chance that the 300\,TeV photon originates from any other source but GRB~221009A, as $p_\text{chance} = 0.9\%$~\cite{Carpet3_300TeV}. We cannot estimate the likelihood of the ``new physics'' or the ``new astrophysics'' needed to explain the EBL and the late arrival puzzles. Hence we cannot compare this chance probability to the case in which these are invoked (that is the case when the photon arises from a different mechanism than the afterglow). However, we can compare this probability to the likelihood of the LIV explanation. Namely, for a given set of LIV parameters, we compare $p_\text{chance}$ with the probability that the observation of the photon at 4310 s is consistent with the LHAASO afterglow light curve and the LIV TOF delay. For example, if one sets $n=1$, the latter probability is $\sim 0.1\%$ (see Fig.~\ref{fig:n-lgEqg}). As this is much smaller than 0.9\%, the possibility of a spurious association is favored. 

\begin{figure*}
    \centering
    \includegraphics[width=\textwidth]{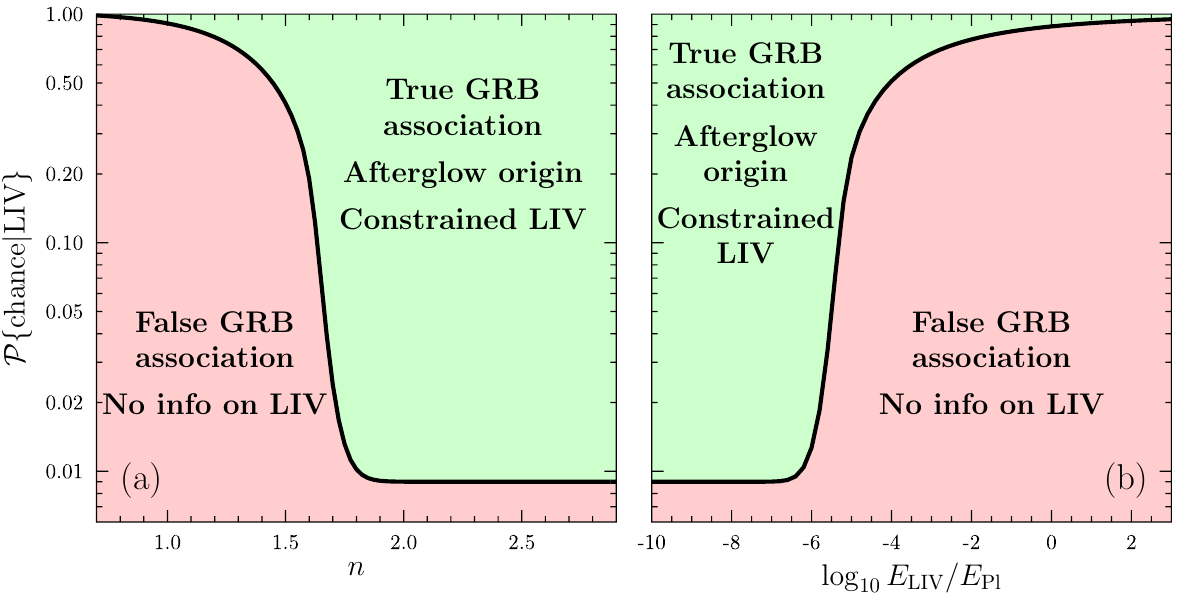}
    \caption{The coincidence probability 
    {conditioned by that LIV has (a) given $n$ or (b) given $\Eqgi{}$,} Eq.~\eqref{eq:coince}, within the assumption that the light curve is the same for the TeV and sub-PeV ranges.}
    \label{fig:coincidence}
\end{figure*}

More generally, the probability for a chance association should be compared to the probability $\mathcal{P}\{\mathrm{consistency|LIV}\}$ that the afterglow light curve is consistent with the TOF delay for given LIV parameters. The latter is the chance that for a given $\Eqg$ and $n$ the absorption constraints from the Carpet-3 photon and the TOF constrains from the TeV LHAASO photons [solid blue (Eq.~\ref{eq:dt_LHAASO}) and black (Eq.~\ref{eq:EBL_Carpet}) lines in Fig.~\ref{fig:n-lgEqg}] are satisfied, while the LIV TOF delay of the Carpet-3 photon is distributed following Eq.~\eqref{eq:prob_LIVTOF}. For fixed $n$ one can write this probability as:
\begin{subequations}
\label{eq:consistency}
\begin{multline}
    \label{eq:consistency:n}
    \mathcal{P}\{\text{consistency}|n\} = \int \diff E_\gamma~ p(E_\gamma) \Biggl[ \mathcal{P}\Bigl\{ \text{LIV TOF} 
    \\
    \left. \left. > \delta t^\text{(LIV)}\left(E_\gamma,\zgrb; \Eqg(n), n \right) \right\}\biggr|^\text{Eq.\eqref{eq:dt_LHAASO}}_\text{Eq.\eqref{eq:EBL_Carpet}} \right],
\end{multline}
where $p(E_\gamma)$ is a normal distribution with a mean $300\,$TeV and a half-width $(38+43)/2\,$TeV and $\Eqg(n)$ marks the boundaries given by the corresponding equations as functions of $n$. Similarly, one can write the consistency probability for fixed $\Eqg$:
\begin{multline}
    \label{eq:consistency:Eqg}
    \mathcal{P}\{\text{consistency}|\Eqg\} = \int \diff E_\gamma~ p(E_\gamma) \Biggl[ \mathcal{P}\Bigl\{ \text{LIV TOF} 
    \\
    \left. \left. > \delta t^\text{(LIV)}\left(E_\gamma,\zgrb; \Eqg, n(\Eqg) \right) \right\}\biggr|^\text{Eq.\eqref{eq:dt_LHAASO}}_\text{Eq.\eqref{eq:EBL_Carpet}} \right].
\end{multline}
\end{subequations}
We use Eqs.~\eqref{eq:dt_LHAASO} and~\eqref{eq:EBL_Carpet} as sharp boundaries, while they should be smeared out according to some distributions. Since the statistical meaning of these constraints is not clear, we put this problem aside. Nevertheless we have checked that possible ways to account for distributions of Eqs.~\eqref{eq:dt_LHAASO} and~\eqref{eq:EBL_Carpet} do not change the qualitative picture below. 

Now we can compare the chance and consistency probabilities in the following way. Suppose that we have only two options: either the 300\,TeV photon is emitted during the same processes as the LHAASO afterglow, or it is not related to this GRB at all. In the first case, the only way to compensate the delay between detections of TeV and sub-PeV photons is to assume that they have different speed, i.e. LIV exists. In the second case, no information about LIV can be gained from the analysis of the Carpet-3 detection. Then for some set of LIV models we can define the conditional chance probability
\begin{multline}
    \label{eq:coince}
    \mathcal{P}\{{\mathrm{chance|LIV}}\} 
    \\
    = \frac{p_\text{chance}}{p_\text{chance} + (1-p_\text{chance})\mathcal{P}\{\text{consistency}|
    {\text{LIV}} 
    \}}  \ .
\end{multline}
This probability is shown in Fig.~\ref{fig:coincidence} for fixed $n$ (a) and fixed $\Eqg$ (b). For any LIV parameter values, there is a non-vanishing chance that the Carpet-3 photon came from the same point in the sky as GRB~221009A just by coincidence. For $n\lesssim 1.4$ or $\Eqg \gtrsim 10^{-4}\Epl$ it is favorable to consider the Carpet-3 photon to be not associated with the GRB, and no constraints on LIV follow. Possible LIV solutions are relevant for $n\gtrsim 1.4$ or $\Eqg \lesssim 10^{-4}\Epl$. Note, however, that as $n$ increases and $\Eqg$ decreases other limits on LIV should be taken into account.

\section{Conclusions}
\label{sec:concl}

The 300 TeV photon observed by Carpet-3 in possible association with GRB~221009A poses two puzzles. 
First, how did such a photon traverse the EBL without being absorbed? Second, why did it arrive at such a late time when no other high energy component was observed?   

The first puzzle surely requires ``new physics.'' Among possible solutions the absorption threshold shift due to LIV can resolve this puzzle. Recently linear LIV with $\Eqgi{1} \sim 25 \Epl$ was proposed as a possible solution \cite{GalantiRoncadelli2025}. This is consistent with current LIV TOF lower limits that are $ 9 \Epl \lesssim \Eqgi{1}$. However, this leaves unresolved the question concerning the late arrival of this photon.

There is no known astrophysical scenario that could explain the emission from GRB 221009A of the Carpet-3 300 TeV photon more than an hour after the LHAASO TeV emission. The late arrival of this photon requires some ``new astrophysics.'' One possibility is a LIV TOF delay due to subluminal propagation of the higher energy photons. But the long delay can be explained neither by $\Eqgi{1} \sim 25\Epl$ nor even by $9\Epl$.  

In this work we have explored the possibility that LIV resolves both puzzles. Namely that in addition to the threshold shift subluminal LIV leads to a delay in the arrival time of the 300 TeV photon relative to LHAASO TeV photons. To do so we begin assuming that the 300 TeV light curve follows the TeV afterglow observed by LHAASO. We then use this light curve to infer the probability of emission of the 300 TeV photon as a function of time. Using this probability we now calculate for given LIV parameters $\Eqg$ and $n$ the likelihood that both the threshold shift and the TOF delay are satisfied. When doing so we take into account the errors in the energy estimate of the Carpet-3 photon and we use the LHAASO modified estimates of the EBL. 

The likelihood for obtaining a LIV solution should be compared with $p_\text{chance}=0.9\%$, the probability for a chance coincidence \cite{Carpet3_300TeV}. Clearly, we should discard LIV solutions whose likelihood is less than $p_\text{chance}$. We find that the resulting acceptable LIV phase space is a range of $\Eqg$ values for each $n$. This range has a lower limit of $n>1.4$ and an upper limit of  $\Eqg<10^{-4} \Epl$. 
Within this range particularly interesting is the $n=2$ model for which we find: $\Eqgi{2} = 1.30_{-0.35(-0.52)}^{+0.56(+3.48)}\times 10^{-7}\Epl$ at 95.4(99.7)\%\ credibility level.
Limits for other integer $n$ values are $\Eqgi{3} = 8.5_{-2.2(-3.4)}^{+2.9(+11.9)}\times 10^{-10}\Epl$, and  $\Eqgi{4} = 6.7_{-1.8(-2.7)}^{+2.1(+6.4)}\times 10^{-11}\Epl$.
These numbers are consistent
\footnote{Limits on subluminal $n=2$ LIV that are much larger than $\Epl$ have been proposed \cite{GalaverniSigl2008} using ultra-high-energy cosmic ray observations. However, those limits depend on a specific interpretation of these observations.}
with existing limits on subluminal non-linear LIV~\cite{SME2009,SatuninEPJC2019,Martinez-Huerta2020}. In particular for quadratic LIV the required $\Eqgi{2}$ range is just above current limits from other gamma-ray sources \cite{Martinez-Huerta2020}.  

GRB 221009A offers a unique opportunity to probe LIV in previously unexplored regimes. The notion that both paradoxes related to the Carpet-3 300 TeV photon might be explained by LIV within acceptable parameter ranges is truly remarkable. Formally, for an acceptable  range of LIV parameters the likelihood of such a solution is larger than the probability for a chance coincidence. In particular, quadratic subluminal LIV with acceptable value of $\Eqgi{2}$ is a viable solution. However, given the theoretical stakes at hand one should hesitate accepting LIV before discarding  other possible explanations and just on the basis of a single event. Still this work points out an intriguing possibility and suggests a way to analyze such future events once they arise.

\begin{acknowledgments}
This research was supported by the ERC advanced grant MultiJets, by the Simons Foundation SCEECS collaboration, and by ISF Grant No. 2126/22.
\end{acknowledgments}

\section*{Data Availability}
The data that support the findings of this article are available from the authors upon reasonable request.

\appendix
\section{The neutron beam model}
\label{sec:app}

The Carpet-3 team \cite{Carpet3_300TeV} suggested the neutron beam  model  as a possible mechanism for the production of the 300 TeV photon. The model, as applied to GRBs \cite{DermerAtoyan2004} involves several stages (i) First, during the prompt GRB phase protons are accelerated to more than PeV in the local fireball frame. They are boosted in the observer frame by the bulk motion of the fireball to EeV. (ii) Interaction of these protons with the surrounding dense photon field produces gamma-rays, electron-positron pairs, neutrinos and neutrons. (iii) The neutrons escape freely. They interact with the surrounding matter in the star forming region surrounding the GRB creating (among other particles) energetic electron-positron pairs. (iv) Those pairs produce the observed TeV photons as synchrotron emission in the surrounding magnetic field. The time delay arises due to the angular spread of the neutron beam \cite{DermerAtoyan2004}. 

This model predicts a copious emission of very high energy neutrino signal. These neutrinos arise at two stages: at stage (ii) when the protons interact with the  photon field producing the neutrons and at stage (iii) when the neutrons interact with the surrounding  matter. The flux of neutrinos should be comparable or larger than the flux of the electron-positron pairs produced at stage (iii) and hence larger than the flux of the emitted synchrotron photons. The strong upper limit on the neutrino flux from GRB 221009A \cite{{Icecube2023,Kruiswijk2023}} poses a serious observational problem. 

At the same time some theoretical questions arise. First, it is not clear at all that a significant flux of $> 1\,$PeV protons are produced during the prompt phase of GRBs. But, even assuming that such protons are produced, within the context of GRBs, the original neutron beam model \cite{DermerAtoyan2004} aims to explain the production of the few hundred MeV emission observed from GRB 941017 [see their equation~(1) and figure~1]. This energy scale is consistent with expected GRB parameters. It is inconsistent with the production of TeV emission. Consider a PeV neutron moving within a jet with a Lorentz factor $\Gamma \lesssim  1000$. This EeV neutron will produce pairs with energy $E_e \lesssim 10^{17}\,$eV. A magnetic field of $\sim 1$ G is needed for these electrons to produce  a 300 TeV synchrotron photon. Such a magnetic field is several orders of magnitude larger than the expected interstellar field, even when considering fields within star forming regions (e.g. \cite{Crutcher2012}). Note that the considerations here were optimal. We assumed  a very high $\Gamma$ during the prompt phase and ignored energy reduction due to the fact that the neutron is moving at an angle to the line of sight (as required to produce the time delay).

\bibliographystyle{apsrev4-2}

\providecommand{\noopsort}[1]{}\providecommand{\singleletter}[1]{#1}%

\end{document}